\documentclass[%
reprint,
superscriptaddress,
showpacs,
amsmath,amssymb,
aps,
]{revtex4-2}
\usepackage{graphicx}
\usepackage{dcolumn}
\usepackage{bm}
\usepackage{CJKutf8}
\usepackage{color}
\usepackage{amssymb}
\usepackage{amsfonts}
\usepackage{palatino}
\usepackage{esint}
\usepackage[colorlinks,urlcolor=blue,linkcolor=blue,citecolor=blue,anchorcolor=blue]{hyperref}
\makeatletter

\newcommand{\Rmnum}[1]{\expandafter\@slowromancap\romannumeral #1@}

\makeatother
\begin{document}

\begin{CJK*}{UTF8}{gbsn}

\preprint{APS/123-QED}

\title{Vortex solitons in disclination quasicrystals}

\author{Hua Zhong}
\affiliation{Key Laboratory for Physical Electronics and Devices, Ministry of Education, School of Electronic Science and Engineering, Xi'an Jiaotong University, Xi'an 710049, China}

\author{Yaroslav V. Kartashov}
\affiliation{Institute of Spectroscopy, Russian Academy of Sciences, Troitsk, Moscow, 108840, Russia} 

\author{Yongdong Li}
\affiliation{Key Laboratory for Physical Electronics and Devices, Ministry of Education, School of Electronic Science and Engineering, Xi'an Jiaotong University, Xi'an 710049, China}

\author{Fangwei Ye}
\email{fangweiye@sjtu.edu.cn}
\affiliation{School of Physics, Chengdu University of Technology, Chengdu 610059, China}
\affiliation{School of Physics and Astronomy, Shanghai Jiao Tong University, Shanghai 200240, China}

\author{Yiqi Zhang}
\email{zhangyiqi@xjtu.edu.cn}
\affiliation{Key Laboratory for Physical Electronics and Devices, Ministry of Education, School of Electronic Science and Engineering, Xi'an Jiaotong University, Xi'an 710049, China}
\affiliation{State Key Laboratory of Human-Machine Hybrid Augmented Intelligence, Institute of Artificial Intelligence and Robotics, Xi'an Jiaotong University, Xi'an 710049, China}

\date{\today}

\begin{abstract}
\noindent
Quasicrystals are ubiquitous in Nature. They represent unique aperiodic materials featuring long-range order that occupy the intermediate niche between exactly periodic and disordered materials. These properties are reflected in unusual evolution dynamics of excitations and unique localization properties of linear eigenmodes supported by quasicrystals. In particular, being structures characterized by discrete rotational symmetry $\mathcal{C}_\nu$ of order $\nu$, photonic quasicrystals are capable to support stable propagation of linear vortex-carrying light beams and vortex solitons. However, the impact of discrete rotational symmetry $\nu$ of quasicrystals on the properties of vortex light states was not investigated so far, as only the systems constructed using the simplest Penrose tiling or corresponding optically induced Penrose lattices were considered in this context. Here we propose a broad class of quasicrystals with global topological defects -- disclinations -- introduced into their structure that allows to produce new quasicrystalline structures with any
desired order of discrete rotational symmetry from basic Penrose structure. Such global topological deformation substantially enriches linear spectrum of quasicrystals, allowing them to support new types of linear vortex states and bifurcating from them families of stable thresholdless vortex solitons with unusual intensity and phase distributions. We found two different classes of stable vortex solitons consisting of in-phase or out-of-phase pairs of closely located bright spots, with total intensity distribution reflecting particular discrete rotational symmetry of the quasicrystal with disclination. Remarkably, even low-charge vortex solitons can be stable in quasicrystals with disclinations, while stability intervals for them broaden with decrease of the discrete rotational symmetry $\mathcal{C}_\nu$ of quasicrystal. Our results expand the theory of localization in quasicrystals to structures with global topological deformation, highlighting new prospects for robust transmission of power or information arising in these systems.
\end{abstract}

\maketitle

\end{CJK*}

\section{Introduction}

Quasicrystals, as aperiodic structures featuring long-range order, occupy a special and very important niche among inhomogeneous materials due to their unique internal structure and symmetry, which are directly manifested in localization and transport properties of excitations. These properties, markedly different for quasicrystals, periodic (crystalline), or disordered inhomogeneous media, stem from qualitatively different localization properties and structure of spectrum of linear eigenstates of these media that are manifested in different physical phenomena. Among the models of quasicrystals is a celebrated aperiodic tiling suggested by Penrose \cite{georgescu.nrp.6.408.2024} that played a central role in the interpretation of scattering experiments in quasicrystalline materials \cite{shechtman.prl.53.1951.1984}. Despite the absence of translational symmetry, quasicrystals nevertheless are characterized by certain discrete rotational symmetry $\mathcal{C}_\nu$ that may substantially affect various physical phenomena encountered in such systems \cite{levine.prl.53.2477.1984, guyot.rrp.54.1373.1991, steurer.aca.74.1.2018}.

As versatile functional materials, quasicrystals have been studied in diverse areas of physics ranging from condensed matter physics~\cite{lifshitz.prl.80.2717.1998, vedmedenko.prl.93.076407.2004, tamura.np.21.974.2025}, physics of matter waves \cite{viebahn.prl.122.110404.2019}, optics~\cite{vardeny.np.7.177.2013, freedman.nature.440.1166.2006, freedman.nm.6.776.2007, levi.science.332.1541.2011, bandres.prx.6.011016.2016, wang.np.18.224.2024, xu.elight.4.9.2024, ivanov.prl.134.113803.2025, zhang.ol.48.2229.2023}, and acoustics~\cite{montero.epl.21.915.1993, han.nc.16.1988.2025}, to name just a few. Aperiodic moir\'e lattices obtained by superposition of two mutually twisted identical sublattices that have attracted considerable attention \cite{wang.nature.577.42.2020, fu.np.14.663.2020, meng.nature.615.231.2023, du.science.379.0014.2023} can also transform into quasicrystals, but only for specific twist angles. However, in contrast to moir\'e lattices that inherit discrete rotational symmetry from their sublattices, quasicrystals (in particular, optically induced ones) may feature any discrete rotational symmetry $\mathcal{C}_\nu$. The remarkable property of both aperiodic moir\'e lattices \cite{wang.nature.577.42.2020} and quasicrystals \cite{wang.np.18.224.2024} is the delocalization-localization transition for their eigenmodes occurring in them upon variation of the properties of the underlying aperiodic potential (in 1D case this effect has been demonstrated in quasiperiodic lattices \cite{lahini.prl.100.013906.2008}). 

Particularly intriguing in this respect is the propagation in aperiodic quasicrystals of vortex-carrying excitations with nonzero orbital angular momentum \cite{shen.light.8.90.2019, chen.ap.7.044001.2025}. Their evolution dynamics is expected to change qualitatively when the quasicrystal is above or below delocalization-localization transition threshold, while discrete rotational symmetry should impose the restrictions on the maximal topological charge of excitations that can be transmitted in stationary manner in $\mathcal{C}_\nu$-symmetric materials, leading to dynamical splitting of phase singularities in states that do not satisfy these restrictions \cite{ferrando.prl.95.043901.2005, kartashov.prl.95.123902.2005, dong.prl.129.123903.2022}. The advantage of optical materials is that their response in addition can be strongly nonlinear, allowing to consider rich interplay between light refraction in inhomogeneous optical potentials and nonlinear self-action. For example, in periodic nonlinear lattices such an interplay leads to the formation of stable vortex lattice solitons \cite{malomed.pre.64.026601.2001, neshev.prl.92.123903.2004, fleischer.prl.92.123904.2004, terhalle.prl.101.013903.2008, terhalle.pra.79.043821.2009}, which however exist only above considerable power threshold (see also reviews \cite{desyatnikov.po.47.291.2005, lederer.pr.463.1.2008, kartashov.rmp.83.247.2011, kartashov.nrp.1.185.2019, mihalache.rrp.69.403.2017, malomed.pd.399.108.2019, pryamikov.jeos.17.23.2021}). The formation of vortex solitons in aperiodic lattices is considerably less studied and requires detailed investigation, as it may create new approaches for control of propagation dynamics and internal structure of light beams with nontrivial phase structure.

While beam localization in nonlinear photonic quasicrystals and the formation of vortex solitons have been considered in Penrose structures with specific fixed discrete rotational symmetry $\mathcal{C}_\nu$ \cite{xie.pre.67.026607.2003, sakaguchi.pre.74.026601.2006, ablowitz.pre.74.035601.2006, law.pra.82.035802.2010, ablowitz.pra.86.033804.2012, antar.jam.2014.848153.2014, sbroscia.prl.125.200604.2020}, there is still no conclusion whether aperiodicity of the material can make such solitons thresholdless and how discrete rotational symmetry of the quasicrystal affects properties and stability of nonlinear vortex-carrying excitations. In particular, all vortex solitons reported so far in Penrose quasicrystals were found to exist only above considerable power thresholds.

\begin{figure*}[t]
	\centering
	\includegraphics[width=\textwidth]{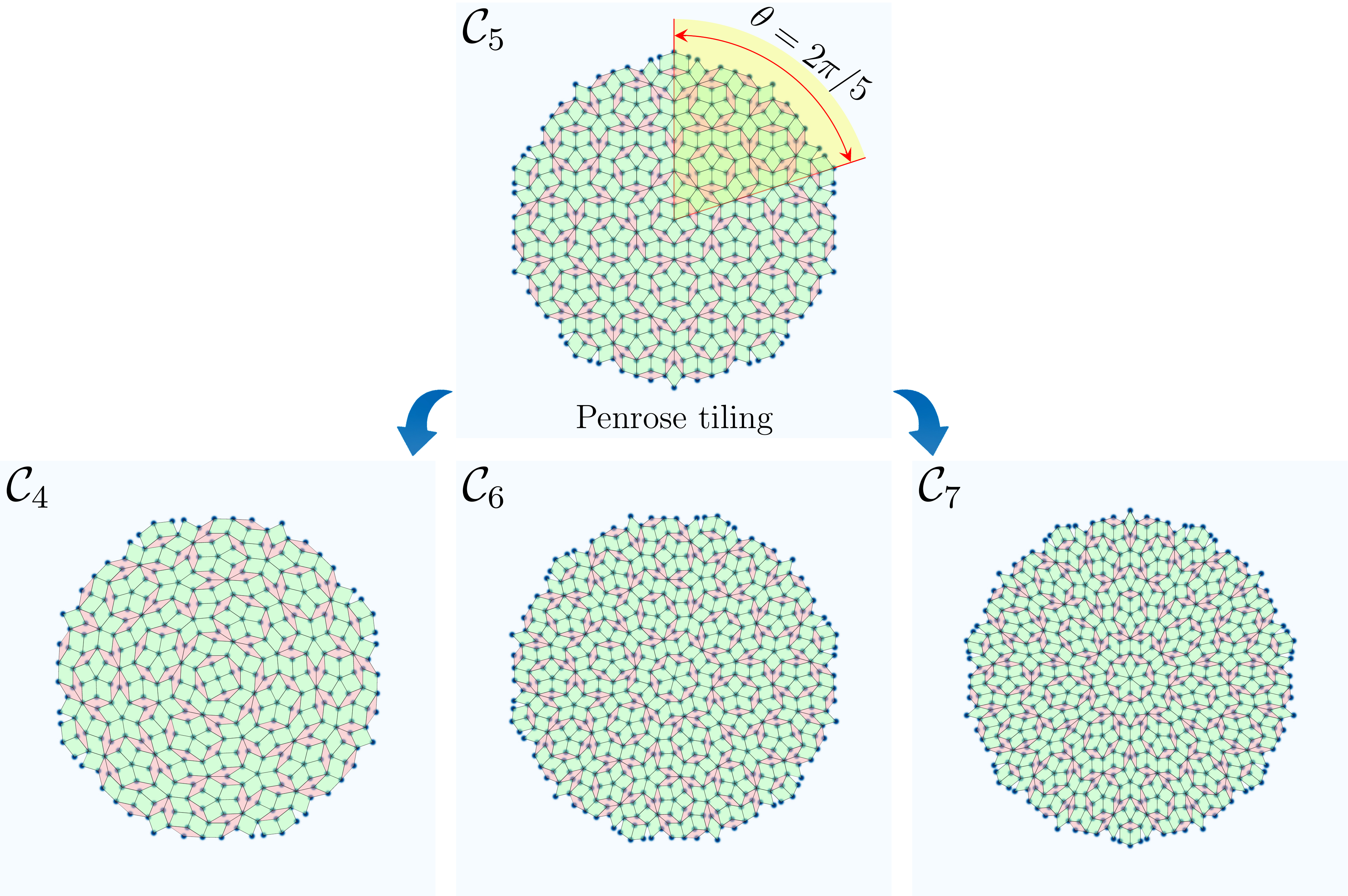}
	\caption{Quasicrystals obtained using Penrose tiling (top) and its disclination counterparts (bottom) with different discrete rotational symmetries $\mathcal{C}_\nu$. The yellow sector with the Frank angle of ${\theta = 2\pi/5}$ can be removed or added into structure to introduce disclination defect.}
	\label{fig1}
\end{figure*}

To answer the fundamental open question about types of vortex solitons that can exist in stable form in quasicrystals with variable discrete rotational symmetry, in this work we have designed a class of quasicrystals with controllable global topological deformation that can be introduced into such structures to continuously alter their discrete rotation symmetry $\mathcal{C}_\nu$. To achieve this, we use the approach that creates topological disclination \cite{harris.sa.237.130.1977, bohsung.prl.58.2277.1987, kleman.rmp.80.61.2008, lin.nrp.5.483.2023} frequently employed in the design of higher-order topological insulators from periodic lattices \cite{peterson.nature.589.376.2021, liu.nature.589.381.2021, ren.apl.8.016101.2023}, but apply it here to aperiodic Penrose quasicrystals. In the frames of this approach, removal or insertion of $\mu$ (here $\mu\le\nu-3$) angular sectors with a Frank angle ${\theta=2\pi/\nu}$ from original $\mathcal{C}_\nu$ quasicrystal lattice and their subsequent filling with waveguides allows to create disclination quasicrystals with new discrete rotational symmetries of the order $\mathcal{C}_{\nu+\mu}$ or $\mathcal{C}_{\nu-\mu}$, overcoming limitations on the position of waveguides imposed by Penrose tiling. 
Notice that while formally for $\mu=3$ one can even construct a $\mathcal{C}_2$ quasicrystal from $\mathcal{C}_5$ one, the former appears too sparse (due to large spacing between waveguides) and does not support vortex states.
The resulting quasicrystals with global topological deformation can support new types of linear vortex states and vortex solitons with symmetries not accessible neither in periodic media, nor in usual quasicrystals. We encountered two different classes of vortex solitons consisting of in-phase and out-of-phase pairs of closely located bright spots, with intensity distribution featuring the same discrete rotational symmetry as underlying quasicrystal. Corresponding \textit{in-phase} and \textit{out-of-phase} vortex solitons bifurcate from proper combinations of linear modes of disclination quasicrystals that are well-localized spatially for sufficiently deep optical potential (above delocalization-localization transition). Since such solitons bifurcate from localized linear eigenstates, they are thresholdless and feature enhanced stability. We show that stability regions are wider in quasicrystals with $\mathcal{C}_{4,5}$ discrete rotational symmetries in comparison with $\mathcal{C}_{6,7}$ quasicrystals. Thus introduction of topological disclinations allows to generate various stable vortex solitons with tailored profiles in aperiodic optical systems.

Our findings are conceptually different from quasicrystalline linear vortex lattices that can be generated with optical induction in free space \cite{jolly.am.22.356.2010, boguslawski.pra.84.013832.2011, li.ol.50.3373.2025}, since we consider evolution of localized vortex inputs in prefabricated quasicrystalline optical potential in nonlinear medium.

\begin{figure*}[t]
	\centering
	\includegraphics[width=\textwidth]{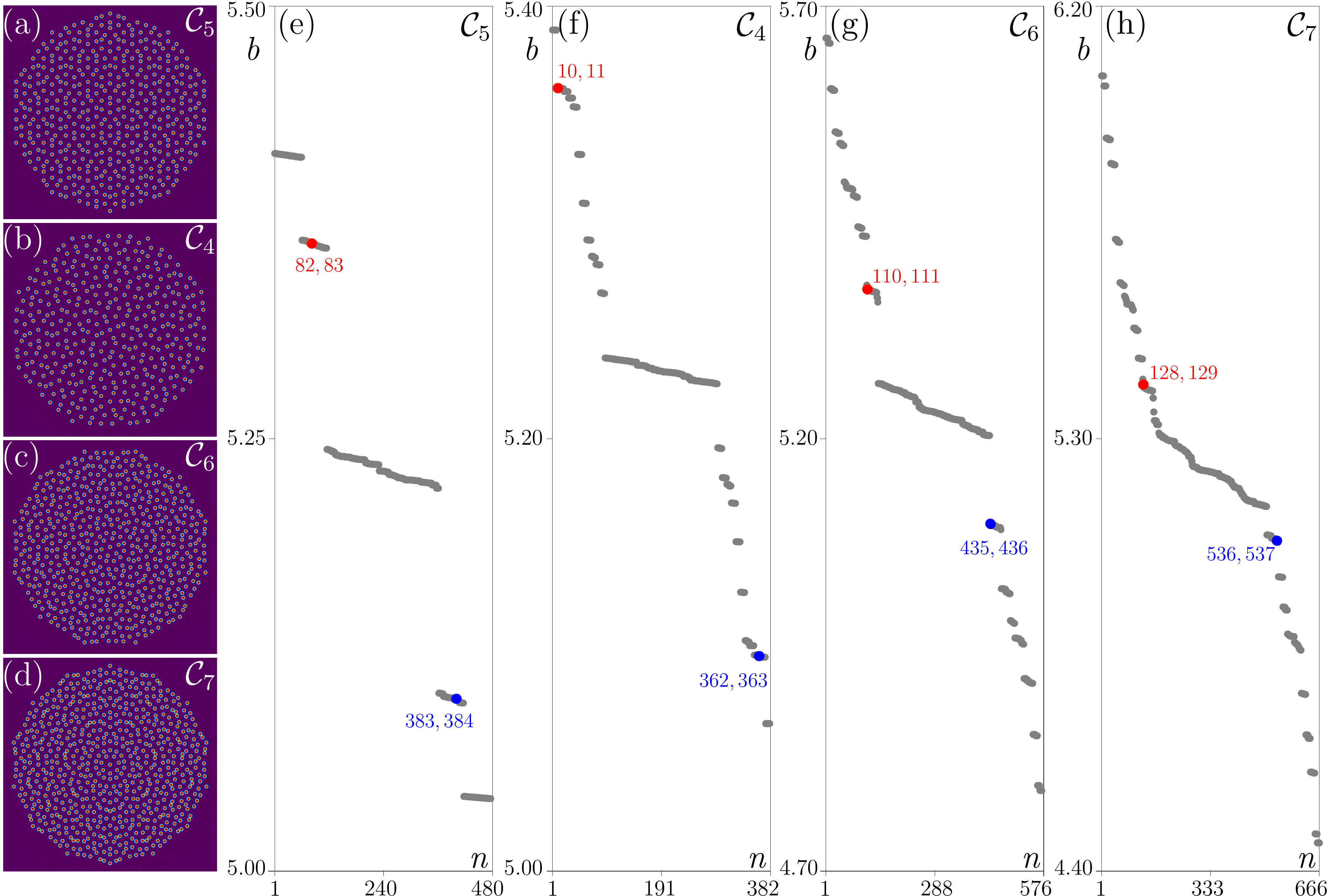}
	\caption{\textbf{Disclination quasicrystalline lattices and their spectra.}
		(a) Quasicrystal generated using Penrose tiling and its linear spectrum in the form of dependence of eigenvalues of all modes $b$ on mode index $n$ (e). (b) $\mathcal{C}_4$ symmetric disclination quasicrystal and its spectrum (f). (c) $\mathcal{C}_6$ symmetric disclination quasicrystal and its spectrum (g). (d) $\mathcal{C}_7$ symmetric disclination quasicrystal and its spectrum (h). Colored dots in (e-h) represent pairs of degenerate states, whose linear combination produces vortices with ${m\pm1}$.
	}
	\label{fig2}
\end{figure*}

\section{Results}

\subsection{Quasicrystals with disclinations}

The propagation dynamics of a light beam along the $z$ axis in the material with focusing nonlinearity and an imprinted quasicrystal refractive index landscape is described by the dimensionless nonlinear Schr\"odinger equation for the amplitude of the light field $\psi$: 
\begin{equation}\label{eq.1}
	i\frac{\partial\psi}{\partial z}=-\frac{1}{2} \nabla^2 \psi-\mathcal{R}\left(x,y\right)\psi-|\psi|^{2}\psi,
\end{equation}
where ${\nabla^2=\partial_x^2 + \partial_y^2}$ is the transverse Laplacian, $x$ and $y$ denote the normalized transverse coordinates, and the function $\mathcal{R}\left(x,y\right)$ describes an optical potential defining quasicrystal that can be constructed from super-Gaussian functions
\begin{equation}\label{eq.2}
	\mathcal{R}(x,y)=p\sum_{m,n}e^{-[(x-x_{m,n})^2+(y-y_{m,n})^2]^2/d^4}
\end{equation}
with $(x_{m,n}, y_{m,n})$ denoting the coordinates of the quasicrystal sites, ${p=12}$ representing the depth of the optical potential, and ${d=0.5}$ being the waveguide width. The Eq.~(\ref{eq.1}) is mathematically analogous to the canonical Schr\"odinger equation in quantum mechanics, with propagation distance $z$ playing the role of the evolution time, and field amplitude $\psi$ being analogous to quantum-mechanical wavefunction. Remarkably, using continuous model (\ref{eq.1}) allows us to account for \textit{all} possible types of couplings in quasicrystals, which is especially important from the point of view of accurate characterization of structure of linear spectrum of the system.

We start by constructing a quasicrystal using famous Penrose tiling of the plane. The sites of such quasicrystal correspond to vertices of tiles of two types, ``thin'' and ``thick'' rhombuses, which are shown with pink and green colors in schematics in Fig.~\ref{fig1}. Throughout this work, we set the edge length of rhombuses to ${a=2.5}$. The resulting aperiodic quasicrystal is characterized by the long-range order, with similar elements appearing after non-equal distances, and with $\mathcal{C}_5$ discrete rotational symmetry.
While we use the scale ${a=2.5}$ to present the results of this work, they remain qualitatively similar for other values of parameter $a$ (to avoid overlap of the waveguides $a$ should exceed approximately $1.5$, as it is required for example in fs-laser writing technology). Decrease of this parameter is accompanied by increase of the coupling strengths between waveguides and overall broadening of linear spectrum, see Fig.~\ref{fig2}. Thus, the eigenvalues of the modes giving rise to vortices shift upon variation of $a$, but they remain in the same parts of the spectrum, i.e. its structure does not change qualitatively.

To go beyond the $\mathcal{C}_5$ symmetry inherent to the Penrose tiling, we introduce disclination defects by removing or adding into above structure $\mu$ sectors with Frank angle ${\theta=2\pi/5}$ (denoted with yellow region in Fig.~\ref{fig1}) and adjust the position of all remaining sites by uniformly stretching or compressing the structure to fill the removed sector or to accommodate the added one. This yields representative quasicrystals with global topological defects having, for example, $\mathcal{C}_4$ or $\mathcal{C}_6$, $\mathcal{C}_7$ discrete rotational symmetry (see bottom row of Fig.~\ref{fig1}) and different central (core) regions. Using this method, one can in principle produce an entire class of disclination quasicrystals with symmetry ranging from $\mathcal{C}_3$ to $\mathcal{C}_\infty$. Further we concentrate on light propagation in quasicrystals with discrete rotational symmetry from $\mathcal{C}_{4}$ to $\mathcal{C}_{7}$. Corresponding refractive index distributions $\mathcal{R}(x,y)$ in these structures are shown in Figs.~\ref{fig2}(a)-\ref{fig2}(d). To limit the number of modes under consideration, we further assume that these structures are sufficiently large, but finite (i.e. 6th-order Penrose tiling was used as a "parent" structure in all cases).
In the \textbf{Supplemental Material}, we display other degenerate states that can form vortex states with higher topological charges.

Such structures can be created using the method of fs-laser writing of waveguides in transparent dielectrics \cite{szameit.jpb.43.163001.2010, mahmood.acs.4.8851.2021, kirsch.np.17.995.2021, arkhipova.sb.68.2017.2023, ren.light.12.194.2023, zhong.light.13.264.2024, kompanets.am.37.2500556.2025, skryabin.prapp.22.064079.2024, zhang.ap.7.034002.2025}, optical induction technique with sets of interfering plane waves in photorefractive crystals~\cite{fleischer.nature.422.147.2003, freedman.nature.440.1166.2006, schwartz.nature.446.52.2007, levi.science.332.1541.2011, chen.ropp.75.086401.2012, wang.nature.577.42.2020, zhong.ap.3.056001.2021} and atomic vapors~\cite{zhang.nc.11.1902.2020, zhang.prl.132.263801.2024}, and various other techniques~\cite{macia.ropp.75.036502.2012, vardeny.np.7.177.2013}. For details of normalizations employed in Eq. (\ref{eq.1}) see Refs.~\cite{kompanets.am.37.2500556.2025,ivanov.prl.134.113803.2025}.

\subsection{Linear spectra and modes of quasicrystals}

To understand how introduction of topological disclination changes the spectrum of quasicrystals it is instructive to consider linear spectra of eigenmodes of corresponding structures. Such eigenmodes are found from Eq.~(\ref{eq.1}) in the form ${\psi (x,y,z) = u(x,y) e^{i bz}}$, where $b$ denotes the propagation constant (eigenvalue) and $u(x,y)$ is the real function describing mode profile. Such profiles satisfy the eigenvalue problem ${bu = (1/2)\nabla^2 u + \mathcal{R} u}$ that can be obtained from Eq.~(\ref{eq.1}) and solved using the plane-wave expansion method.

\begin{figure*}[t]
	\centering
	\includegraphics[width=\textwidth]{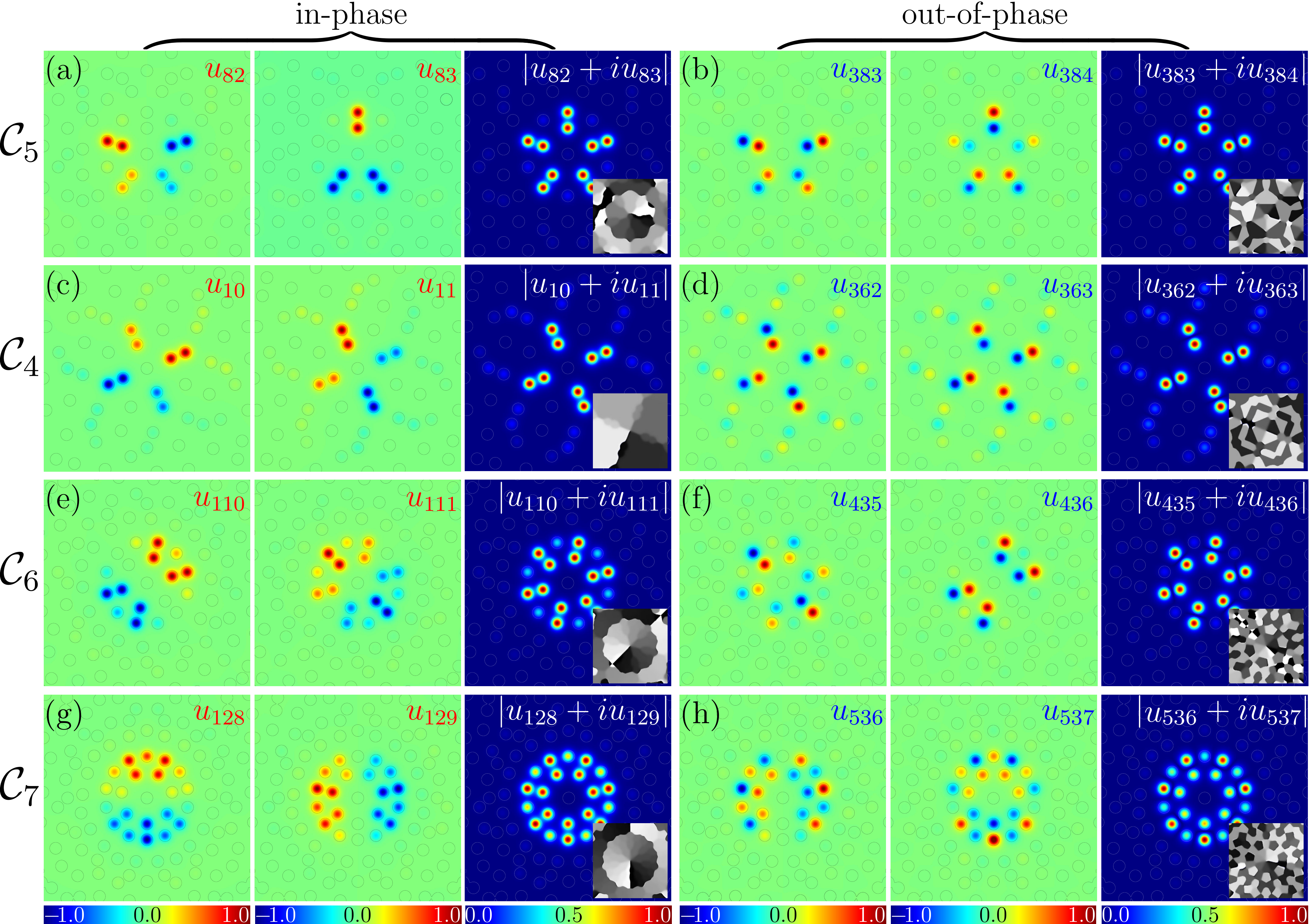}
	\caption{\textbf{Vortex state with ${m=\pm1}$ composed of degenerated modes.}
		(a) Profiles of two degenerate modes $u_{82}$ and $u_{82}$, and modulus and phase (inset) distributions in in-phase vortex state ${u_{82}+iu_{83}}$ constructed using these modes in $\mathcal{C}_5$ quasicrystal constructed using Penrose tiling. (b) Degenerate modes $u_{383}$ and $u_{384}$ and out-of-phase vortex supported by $\mathcal{C}_5$ quasicrystal. Other panels show profiles of degenerate modes and in-phase and out-of-phase vortices constructed on them in $\mathcal{C}_4$ (c),(d), $\mathcal{C}_6$ (e),(f), and $\mathcal{C}_7$ (g),(h) disclination quasicrystals. For convenience of presentation and comparison of different modes, the peak amplitude of each linear mode in this figure is normalized to 1. The faint circles in each panel represent the lattice sites. Profiles are shown within the window ${-10\leq x,y \leq10}$.
	}
	\label{fig3}
\end{figure*}

Linear spectrum of ``parent" $\mathcal{C}_5$ Penrose quasicrystal with selected tiling in the form of dependence of propagation constant $b=b_n$ (sorted such that modes with lower indices have higher propagation constants) on mode index $n$ is presented in Fig.~\ref{fig2}(e). One can see that propagation constants are organized in five mini-bands. Among them, the modes with indices $n=82,83$ (red dots) and $n=383,384$ (blue dots) are localized in the very center of quasicrystal and form degenerate pairs. Profiles of modes ${n=82,83}$ are depicted in Fig.~\ref{fig3}(a), whereas those of modes ${n=383,384}$ are shown in Fig.~\ref{fig3}(b). Closely located pairs of spots in profiles of these modes are respectively in-phase and out-of-phase. Coexistence of two such types of localized modes close to the center of quasicrystal is a result of chosen Penrose tiling. Linear combinations of these modes $u_{82}\pm iu_{83}$ and $u_{383}\pm iu_{384}$ produce vortex-carrying states with topological charges ${m=\pm1}$, whose modulus and phase distributions are depicted in Figs.~\ref{fig3}(a) and \ref{fig3}(b). Notice distinctly different phases of these states: phase of the vortex produced by out-of-phase states is strongly radially modulated in contrast to phase of the vortex produced by in-phase states. So-constructed \textit{in-phase} and \textit{out-of-phase} vortices can propagate without changes in their shapes in $\mathcal{C}_5$ quasicrystal and, to the best of our knowledge, they have never been reported before. Among surprising properties of vortex states is their pronounced localization despite the fact that their propagation constants are located in spectral mini-bands. This is a consequence of sufficiently large depth $p$ of waveguides considered here, that ensures that we work in the regime well above delocalization-localization transition in $\mathcal{C}_5$ quasicrystal. According to our simulations, the degenerate localized modes that would have equal propagation constants and that would be able to produce vortical states localized at the off-center positions in quasicrystal lattices are absent in linear spectrum.

Introduction of global deformation that results in the formation of $\mathcal{C}_4$ quasicrystal with disclination results in substantial modification of the linear spectrum [see Fig.~\ref{fig2}(f)], where one observes the formation of multiple mini-bands with close propagation constants. This is because the introduction of disclination changes inter-waveguide spacing differently at different distances from the center of quasicrystal, and instead of only two characteristic nearest-neighbor spacing values (as in $\mathcal{C}_5$ quasicrystal) one now obtains the structure with multiple nearest-neighbor spacings. Still, two pairs of degenerate modes localized near the center of $\mathcal{C}_4$ quasicrystal are found in linear spectrum in Fig.~\ref{fig2}(f), whose linear combinations produce in-phase and out-of-phase vortex states depicted in Figs.~\ref{fig3}(c) and \ref{fig3}(d). Linear spectra of $\mathcal{C}_6$ and $\mathcal{C}_7$ disclination quasicrystals displayed in Figs.~\ref{fig2}(g) and \ref{fig2}(h), respectively, show similar fractionalization, but in all cases it is possible to construct stationary vortex-carrying states residing in the centers of corresponding quasicrystals [see modulus and phase distributions in Figs.~\ref{fig3}(e)-\ref{fig3}(h)] that feature the same discrete rotational symmetry as that of the quasicrystal. The introduction of disclination therefore appears as a powerful tool allowing the construction of spatially localized vortex states with a desired symmetry. It should be stressed that while we concentrate here on linear combinations producing ${m=\pm 1}$ vortices, there exist other groups of degenerate modes in linear spectrum of the system that allow the construction of vortices with higher charges permitted by the discrete rotational symmetry of the quasicrystal \cite{ferrando.prl.95.043901.2005, kartashov.prl.95.123902.2005}. Thus $\mathcal{C}_5$ and $\mathcal{C}_6$ quasicrystals with disclinations allow construction of vortex states with topological charges up to $m=\pm2$, while $\mathcal{C}_7$ quasicrystals allow construction of vortex states with charges up to ${m=\pm3}$, which are displayed in \textbf{Supplemental Material}. Note that the maximal value of the topological charge of the vortex that can be supported by the quasicrystal with disclination can be predicted using the arguments of group theory, see details in \cite{ferrando.prl.95.043901.2005, kartashov.prl.95.123902.2005}. Thus, in $\mathcal{C}_4$ structures only the topological charges ${m=\pm1}$ are compatible with discrete rotational symmetry of the lattice, while vortex states with ${|m|\ge2}$ are forbidden as irreducible representations of the $\mathcal{C}_4$ group. Linear spectrum of corresponding quasicrystals directly reflects this property, since discrete rotational symmetry $\mathcal{C}_k$ determines the number of pairs of localized degenerate states (producing vortices) that can be found in the entire spectrum. In each case, for a given charge $m$ it is possible to construct both in-phase and out-of-phase types of vortices (by analogy with results shown in Fig.~\ref{fig3}), by using combination of proper linear modes. Furthermore, as the order of the discrete rotational symmetry increases, the formation of vortex states with even higher charges can be expected. Therefore, the ability to alter discrete rotational symmetry renders disclination quasicrystals a versatile platform for the generation of vortices with tailored charges that remain spatially localized in these aperiodic systems above delocalization-localization transition (above certain threshold value of $p$).

\subsection{Vortex solitons and stability analysis}\label{stable}

In this section we obtain vortex soliton families bifurcating from the above linear localized vortex states. The profiles of such vortex solitons can be found from the equation ${bu = (1/2)\nabla^2 u + \mathcal{R} u+ |u|^2u}$, where we account for focusing nonlinearity of the material, and where the function $u$ describing soliton profile is now complex and determined by the propagation constant $b$, which is now a free parameter. We solved this equation using the Newton method. We primarily concentrate here on vortex soliton families originating from out-of-phase linear vortex states, but their in-phase counterparts can be analyzed using similar approach.

\begin{figure*}[t]
	\centering
	\includegraphics[width=\textwidth]{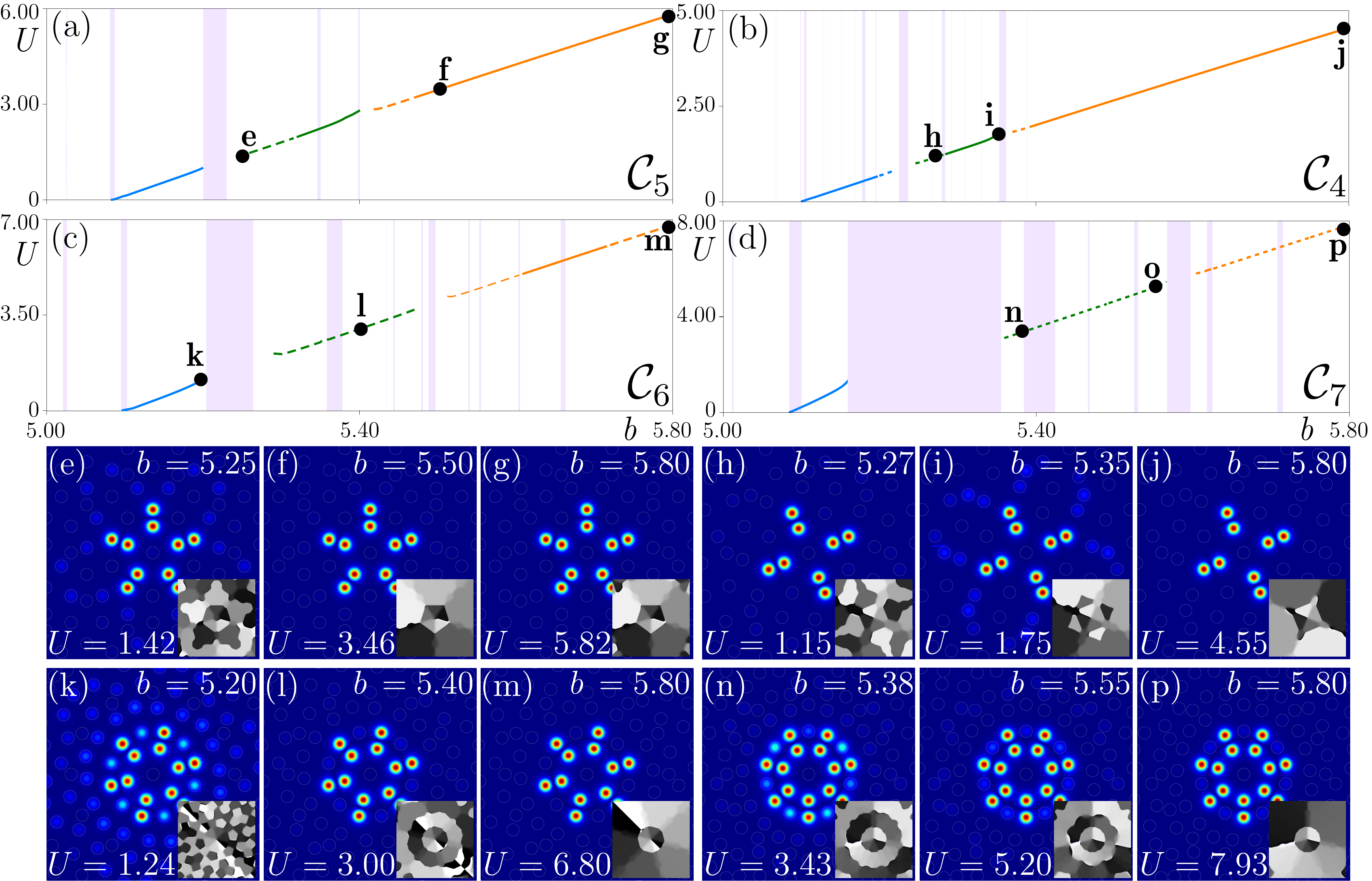}
	\caption{\textbf{Vortex soliton families with ${m=\pm1}$.}
		(a-d) $U(b)$ dependencies for out-of-phase vortex soliton families with topological charge $|m|=1$ in quasicrystals with different discrete rotational symmetries. Gray areas correspond to mini-bands from linear spectra in Fig. \ref{fig2}. (e-p) Field modulus and corresponding phase (insets) distributions of the vortex solitons corresponding to the dots with letters in $U(b)$ dependencies.
	}
	\label{fig4}
\end{figure*}

Out-of-phase vortex soliton families with ${|m|=1}$ in Penrose quasicrystal with $\mathcal{C}_5$ discrete rotational symmetry are presented in Fig.~\ref{fig4}(a) by the dependence of soliton power ${U = \iint |u|^2 dxdy}$ on the propagation constant $b$. Because nonlinearity is focusing, bifurcation from linear vortex occurs in the direction of increasing $b$. When soliton propagation constant $b$ approaches the eigenvalue $b_{n=383,384}$ of degenerate linear modes giving rise to out-of-phase vortex, both amplitude $\textrm{max}|u|$ and power $U$ of soliton vanish, i.e. such states are thresholdless for our parameters [see blue line in Fig.~\ref{fig4}(a)] in contrast to all previously reported vortex solitons in quasicrystals. When propagation constant $b$ of vortex soliton crosses mini-bands [represented by gray regions in Fig.~\ref{fig4}(a)], the coupling with different modes belonging to this mini-band unavoidably occurs and this leads to substantial shape transformation and overall expansion of vortex soliton (because modes from mini-bands can be concentrated on different spatial locations inside quasicrystal). Because coupling may occur with multiple modes that renders $U(b)$ dependence very complex and yields coexistence of many different types of vortical solutions, we do not trace it inside mini-bands. The family of well-localized out-of-phase vortex solitons residing in the center of quasicrystal can be continued in other spectral gaps, both finite (green line) and semi-infinite (orange line) ones, as shown in Fig.~\ref{fig4}(a). 
Notice that the internal structure of vortex soliton with representative $\pi$ phase difference between spots in close pairs is maintained also in higher gaps, but overall phase distribution gradually simplifies [the number of phase jumps in radial direction decreases in higher gaps as one can see from the insets in Fig.~\ref{fig4}(e) with soliton profiles]. Thus, nonlinearity provides a powerful knob to control localization, internal and phase structure of vortex excitations in aperiodic materials.

\begin{figure*}[t]
	\centering
	\includegraphics[width=\textwidth]{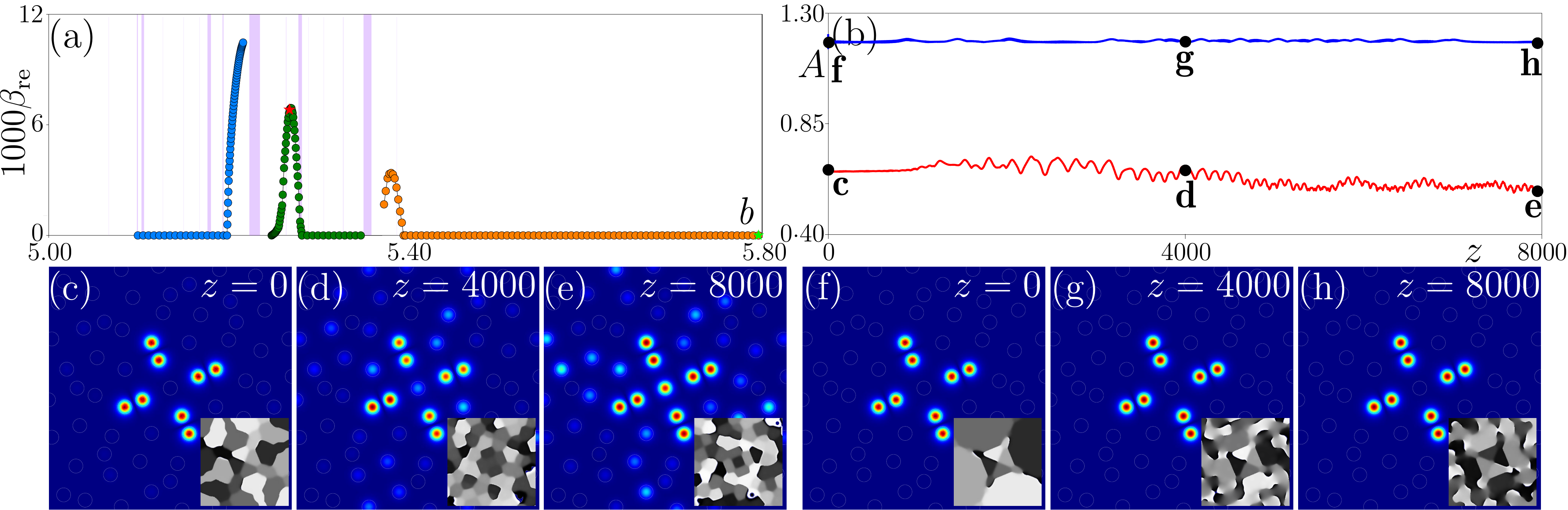}
	\caption{\textbf{Linear stability analysis of vortex solitons in the $\mathcal{C}_4$ quasicrystal.}
		(a) Perturbation growth rate $\beta_{\rm{re}}$ versus propagation constant $b$ for out-of-phase vortex solitons in $\mathcal{C}_4$ quasicrystal. (b) Peak amplitude $A=\textrm{max}|\psi|$ of representative stable (blue curve) and unstable (red curve) perturbed vortex solitons versus propagation distance. (c-h) Field modulus phase distributions of solitons at different propagation distances corresponding to the dots in panel (b).}
	\label{fig5}
\end{figure*}

The families of the out-of-phase vortex solitons in $\mathcal{C}_4$, $\mathcal{C}_6$, and $\mathcal{C}_7$ quasicrystals with disclinations are presented in Figs.~\ref{fig4}(b)-\ref{fig4}(d), respectively. Because the complexity of linear spectrum substantially increases upon introduction of disclination into quasicrystal, resulting in the appearance of multiple mini-bands in spectrum (in comparison with only five mini-bands in Penrose $\mathcal{C}_5$ case), vortex solitons cross such multiple mini-bands upon increase of propagation constant $b$. Still, coupling occurs only with selected small subset of mini-bands (since for some mini-bands it is prohibited by different symmetry, or by very different locations of the vortex soliton and linear states from this mini-band). As a result, it was possible to trace $U(b)$ families in wide power ranges, as shown in Figs.~\ref{fig4}(b)-\ref{fig4}(d). It should be mentioned that in real experimental conditions, when vortex states are excited with localized inputs with designed phase distribution, one would observe namely such transitions between soliton families from different spectral gaps upon increase of input power (provided that corresponding solitons are stable), perhaps accompanied by spreading in narrow power intervals, where coupling with other modes can occur. Representative field modulus and phase distributions in out-of-phase vortex solitons in $\mathcal{C}_4$, $\mathcal{C}_6$, and $\mathcal{C}_7$ quasicrystals are depicted in Figs.~\ref{fig4}(h)–\ref{fig4}(p). One can see how discrete rotational symmetry of intensity distribution of the vortex soliton increases with increase of lattice symmetry. In some profiles taken sufficiently close to mini-bands [see dots in corresponding $U(b)$ dependencies] one can see broadening of soliton due to coupling with linear modes in other spatial locations. Just as in $\mathcal{C}_5$ case the tendency to simplification of phase distribution is visible in higher spectral gaps.

Stability of solitons represents one of the most fundamental and crucial prerequisites for their potential experimental observation. We use two-fold approach to characterization of stability of the obtained families: linear stability analysis and direct propagation of the perturbed states allowing to assess stability beyond perturbative approach, even when actual perturbations are not small. In the latter case, we add random noise into input vortex solitons with amplitude up to $10\%$ of soliton’s peak amplitude, and propagate them over long distance ${z\sim8000}$. A perturbed soliton is considered stable if it retains its integrity and characteristic phase structure throughout propagation; otherwise, it is classified as unstable. According to this method, the largest part of the out-of-phase vortex soliton families in $\mathcal{C}_4$ and $\mathcal{C}_5$ quasicrystals corresponding to solid lines in Figs.~\ref{fig4}(a) and \ref{fig4}(b) are stable. Instabilities are encountered only in narrow ranges of $b$ values [see dashed lines in Figs.~\ref{fig4}(a) and \ref{fig4}(b)], typically close to mini-bands. Notably, as discrete rotational symmetry order of the system increases, the instability domains broaden, see dashed parts of soliton families in Figs.~\ref{fig4}(c) and \ref{fig4}(d) corresponding to $\mathcal{C}_6$ and $\mathcal{C}_7$ quasicrystals. In all cases, vortex solitons are stable near bifurcation points from linear vortex modes.

\begin{figure*}[t]
	\centering
	\includegraphics[width=\textwidth]{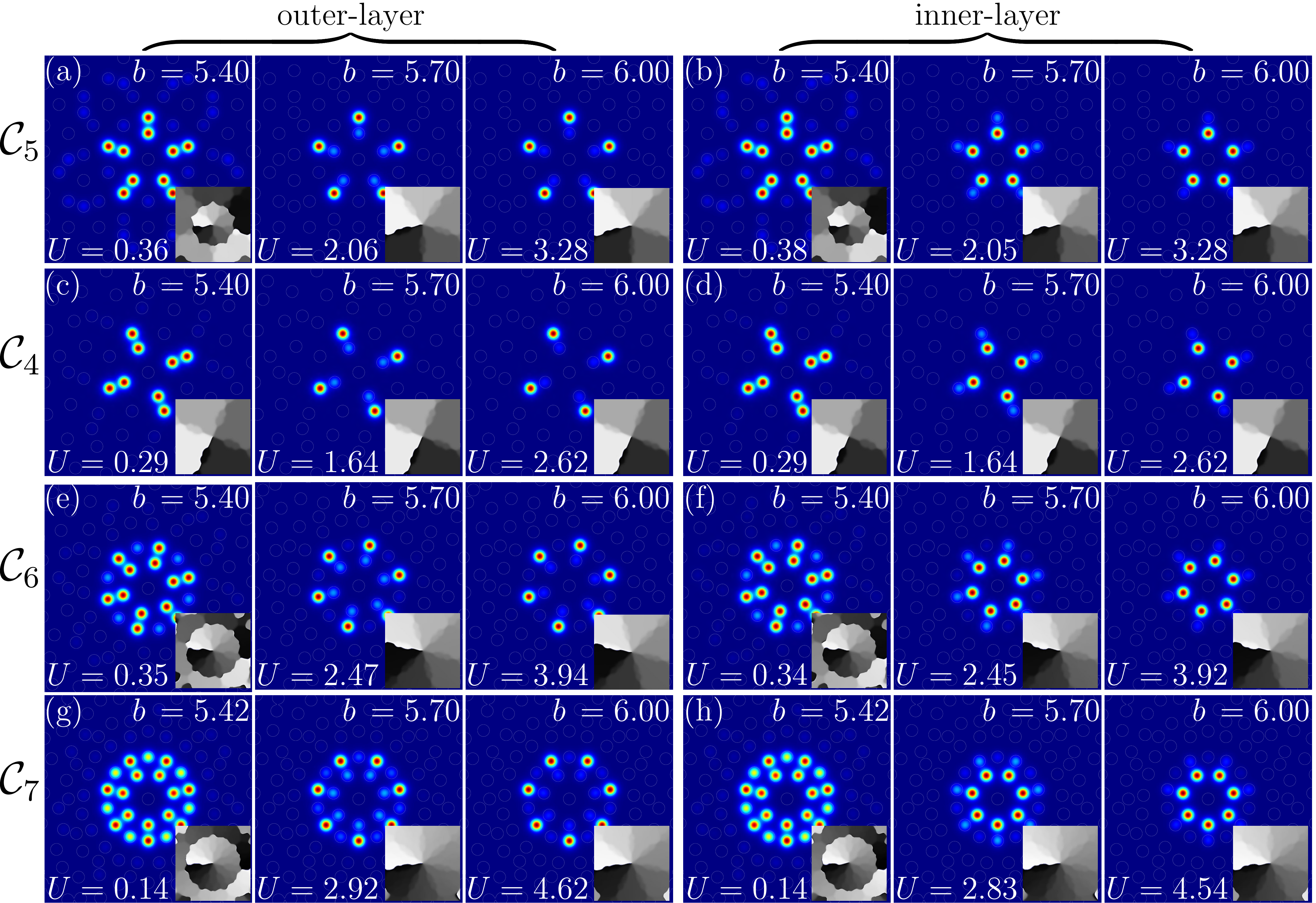}
	\caption{Field modulus and phase (insets) distributions for ${m=\pm1}$ in-phase vortex solitons illustrating their rich shape transformations with increase of propagation constant $b$ in (a,b) original Penrose quasicrystal with $\mathcal{C}_5$ symmetry, and quasicrystals with disclinations with (c,d) $\mathcal{C}_4$, (e,f) $\mathcal{C}_6$, and (g,h) $\mathcal{C}_7$ symmetry. Parameters of quasicrystals are the same as in Fig.~\ref{fig2}.}
	\label{fig6}
\end{figure*}

To complete the above analysis, we performed also linear stability analysis that requires superimposing small perturbations $w$ and $v$ onto the vortex soliton profile $u$ as follows
\begin{equation}\label{eq.5}
	\psi = [ u(x, y)+v(x, y) e^{\beta z}+w^{*}(x, y) e^{\beta^{*} z} ] e^{i b z}
\end{equation}
where ${v,w\ll u}$ , while $\beta=\beta_\textrm{re}+i\beta_\textrm{im}$ is the complex perturbation growth rate. Substituting Eq.~(\ref{eq.5}) into Eq.~(\ref{eq.1}) and linearizing the resulting equation around the stationary solution $u$, one arrives to the following linear eigenvalue problem:
\begin{equation}\label{eq.6}
	\begin{split}
		i \beta v= \,&-\frac{1}{2} \nabla^2 v-({\mathcal R}-b) v-2|u|^{2} v-|u|^{2} w,\\
		i \beta w = \,& +\frac{1}{2} \nabla^2 w + ({\mathcal R}-b) w + 2|u|^{2} w + |u|^{2} v.\\
	\end{split}	
\end{equation}
Solving the problem (\ref{eq.6}) using standard eigenvalue solver, we obtain the dependence of the perturbation growth rate $\beta$ (and associated perturbation profiles $v,w$) on the propagation constant $b$ for a given family of vortex solitons. A vortex soliton $u$ is deemed linearly stable if the real part of the perturbation growth rate ${\beta_{\rm{re}} \le 0}$ for all possible perturbations; in this case, the soliton will only exhibit small-amplitude oscillations during propagation even in the presence of small perturbations. Conversely, if at least one perturbation mode yields ${\beta_{\rm{re}} > 0}$, the vortex soliton is unstable and will decay as it propagates. In Fig.~\ref{fig5}(a), we show maximal real part of perturbation growth rate $\beta_{\rm re}(b)$ for the vortex soliton families in the $\mathcal{C}_4$ quasicrystal, as a representative example. Notably, the results from linear stability analysis are in excellent agreement with those obtained via direct propagation of perturbed states [cf.~dashed lines in Fig.~\ref{fig4}(b)]. Red stars in Fig.~\ref{fig5}(a) correspond to, respectively, unstable and stable states h and i in Fig.~\ref{fig4}(b). Peak amplitude of these perturbed states during propagation is shown in Fig.~\ref{fig5}(b), while field modulus and phase distributions at different distances are shown in Figs.~\ref{fig5}(c)-\ref{fig5}(e). The unstable state looses its vorticity and shows considerable amplitude oscillations upon propagation due to tunneling of power into other sites of quasicrystal (red curve), while stable state keeps its internal structure and shows minimal oscillations (blue curve).

As it was mentioned above, disclination quasicrystals also support rich families of in-phase vortex solitons. These solitons demonstrate very interesting nonlinear transformations upon increase of power, when different branches of solutions can bifurcate from a given family (bifurcating from linear in-phase vortex state), characterized by equal intensities of spots in close pairs (see left outermost solutions in Fig.~\ref{fig6}). With increase of the propagation constant $b$ two branches of solutions emerge in one of which light concentrates in outer sites from close pairs, as in Figs.~\ref{fig6}(a,c,e,g), while in solutions belonging to other branch, light concentrates in inner sites from close pairs, as in Figs.~\ref{fig6}(b,d,f,h). This is a result of classical symmetry-breaking bifurcation in focusing medium that actually doubles the number of possible in-phase solutions in a given spectral gap [we do not show here corresponding $U(b)$ dependencies due to their complexity]. Such bifurcations occur for all types of quasicrystals considered here.
Just like in out-of-phase solitons, the intensity distributions of in-phase solitons reflect discrete rotational symmetry of corresponding quasicrystal. Co-existence of two different types of in-phase and out-of-phase vortex solitons in quasicrystals is a result of their rich and aperiodic spatial structure determined by the ``parent'' Penrose tiling.

In the above analysis we concentrated only on vortex soliton families with topological charge $|m|=1$. In \textbf{Supplemental Material} we provide other examples of the degenerate eigenmodes, whose linear combinations can produce vortices with higher topological charges $|m|>1$ (in particular, in $\mathcal{C}_6$ and $\mathcal{C}_7$ quasicrystals). All such linear vortex modes give rise to the families of thresholdless vortex solitons. Our simulations show that vortex solitons with different topological charges emerging from different linear modes may coexist within some gaps. Still, vortex solitons with different topological charges do not couple because they are orthogonal (due to different charges), i.e. their overlap integral is zero even in nonlinear case.

\section{Discussions}
\subsection{Approximations and limitations}
Before closing, we would like to stress that the approach to creation of quasicrystals with disclinations suggested here is powerful and leads to multiple unusual thresholdless vortex states, but it is, of course, not free from certain limitations.

First, one has to take into account that increasing discrete rotational symmetry of the quasicrystal structure due to addition of sectors with larger and larger Franck angles leads to gradual increase of the density of waveguides and sooner or later they will start overlapping. This may qualitatively affect linear spectrum of the system and, besides possible technological difficulties in creation of such structures with tightly packed waveguides, this imposes natural limitations on the maximal order of discrete rotational symmetry that can be realized with this approach. Likewise, removing too large Franck angle will result in quasicrystal structure with too sparse arrangement of practically non-interacting waveguides (that would be reflected in a very narrow spectral bands).

Second limitation is connected with the fact that the original $\mathcal{C}_5$ structure produced by Penrose tiling of the plane that we use as a platform for construction of disclination arrays is not unique. Hence, the selection of other tiling may produce disclination quasicrystals with different spectral features. They will still support vortex states with topological charges predicted by our analysis, but the details of the modal shapes may be different, leading to different nonlinear families and bifurcations.

\begin{figure*}[t]
\centering
\includegraphics[width=\textwidth]{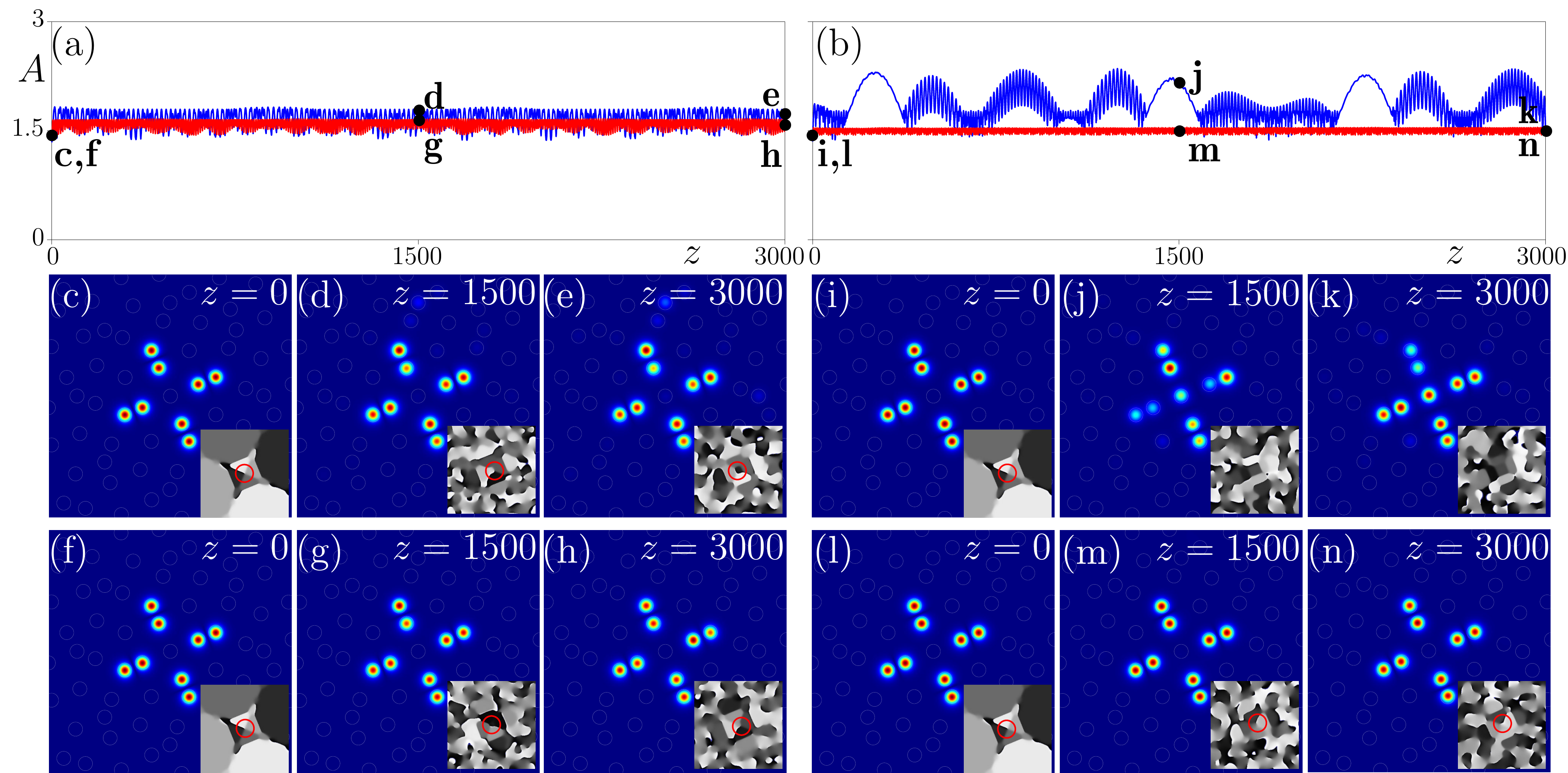}
\caption{Peak amplitude of the vortex state upon propagation in $\mathcal{C}_4$ quasicrystal with $3\%$ (a) and $6 \%$ (b) positional and depth disorder. Blue and red curves correspond to linear propagation and nonlinear propagation, respectively. (c-h) Field modulus and phase distributions corresponding to dots in (a). (i-n) Field modulus and phase distributions corresponding to dots in (b). Red circles highlight phase singularity in the center of the pattern.}
\label{fig7}
\end{figure*}

Finally, it should be mentioned that our results and methods are applicable for structures with shallow transverse refractive index modulation and for paraxial light beams. At the same time, these ideas can be potentially extended to non-paraxial models and nanoscale quasicrystalline devices using the full-wave analysis~\cite{hwang.np.18.286.2024}.

\subsection{Influence of disorder on vortex states in quasicrystals}

Robustness of vortex solitons in quasicrystals with respect to perturbations of initial waveforms $\psi$ has been studied by us in Section~\ref{stable}. However, under realistic experimental conditions, certain level of disorder/imperfections may present in the underlying optical potential landscape $\mathcal{R}$ breakings its ideal $\mathcal{C}_\nu$ discrete rotational symmetry. In this section we consider the impact of such disorder on propagation dynamics of vortex states in both linear and nonlinear regimes. To this end, we introduce both diagonal (depth) and off-diagonal (position) disorder into underlying lattice (\ref{eq.2}) by letting depths of individual waveguides in the lattice to fluctuate (uniform distribution) within the range $[p(1-\delta_p),p(1+\delta_p)]$, while deviations of positions of waveguide centers $(x_{m,n},y_{m,n})$ from their regular values are allowed to vary within the interval $[a(1-\delta_a),a(1+\delta_a)]$. We gradually increase the strength of disorder characterized by parameters $\delta_p,\delta_a$ taking for simplicity $\delta_p=\delta_a$ and consider its impact on the evolving vortex modes in linear and nonlinear regimes. We considered different realizations of disorder with strength up to $10\%$. Modern fs-laser writing technology usually allows to achieve much better accuracy of positioning (at the level of several nanometers for typical spacing of tens of micrometers) and uniformity of waveguide depths across the structure (especially when multi-track writing technology is employed~\cite{skryabin.prapp.22.064079.2024}). Typical scenarios of propagation of vortices in exemplary $\mathcal{C}_4$ structure with disorder are presented in Fig.~\ref{fig7}. The presence of weak disorder ($3\%$ in this case) in underlying structure causes oscillations of peak amplitude $A$ of vortex state during propagation that are stronger in linear case (blue curve) in comparison with nonlinear case (red curve) [Fig.~\ref{fig7}(a)]. While disorder causes slight shape transformations of the input state and may cause slight displacement of phase singularity from the center of the lattice [see representative modulus and phase (insets) distributions in Figs.~\ref{fig7}(c)-\ref{fig7}(e) for linear case, and in Figs.~\ref{fig7}(f)-\ref{fig7}(h) for nonlinear case], the main outcome of the analysis is that vortex withstands the presence of weak disorder in the underlying lattice. This conclusion holds for disorder levels up to approximately $4\%$. When the strength of disorder is increased to approximately $6\%$ multiple disorder realizations appear where the initial vortex-carrying state exhibits considerable shape transformations in linear case (the spots in vortex profile change their intensities, coupling with other modes is possible that leads to overall expansion of the propagating field, and vorticity in the center of the structure is lost), the example is shown in Figs.~\ref{fig7}(i)-\ref{fig7}(k). Such transformations are usually accompanied by considerable amplitude oscillations [Fig.~\ref{fig7}(b)]. However, when the same input vortex state propagates in nonlinear medium, nonlinearity may play a strong stabilizing action, even in the presence of disorder. In this case, the amplitudes of spots in vortex profile become nearly equal in amplitude, the amplitude oscillations are suppressed, and vorticity is clearly preserved, as shown in Figs.~\ref{fig7}(l)-\ref{fig7}(n). Such behavior has been observed for the majority of disorder realizations. These results confirm that vortex modes are experimentally observable in quasicrystals with disclinations.

\section{Conclusion}

Summarizing, we have investigated the formation of vortex solitons in quasicrystals with introduced into them global topological deformation resulting in formation of disclination structures with variable discrete rotational symmetry. Starting from ``parent" $\mathcal{C}_5$ Penrose tiling, we obtained various aperiodic disclination lattices with $\mathcal{C}_{4,6,7}$ discrete rotational symmetry. Despite considerable fractionalization of spectrum due to introduction of disclination, each of these quasicrystals support sets of localized (above delocalization-localization transition threshold) degenerate states in the center of the structure that can produce in-phase and out-of-phase vortices with topological charges limited by the discrete rotational symmetry of the underlying structure. We also encountered rich families of in-phase and out-of-phase vortex solitons bifurcating from such linear vortical states and existing across wide range of powers. The localization, intensity and phase structure of such states is controlled by their power in nonlinear medium, thereby illustrating that nonlinearity may serve as a convenient tool for control of structure of propagating light beams in such aperiodic materials with long-range order. Our results may be useful for the design of new switching and transmission devices for states carrying nonzero orbital angular momentum, as well as new classes of lasers and quantum memory harnessing new types of discrete rotational symmetries of aperiodic refractive index landscapes.

The results obtained here can be also generalized to the case of quasicrystals with defocusing nonlinearity following the method utilized in this work. Such states would bifurcate from linear vortex modes in the direction of decreasing propagation constants. Preliminary analysis shows that vortex solitons in defocusing medium demonstrate exceptional stability and resilience to disorder.

\begin{acknowledgments}		
This work was supported by the National Natural Science Foundation of China (Grant Nos. 12304370 and 12474337), 
the Natural Science Basic Research Program of Shaanxi Province (Grant Nos.~2024JC-JCQN-06, 2025JC-QYCX-006, 2025CY-YBXM-037), 
the Postdoctoral Research Project of Shaanxi Province (Grant No. 2023BSHYDZZ14), 
the Sichuan Science and Technology Program (Grant No. 2025ZNSFSC1458), 
the Fundamental Research Funds for the Central Universities (Grant No. xzy012024135), 
the Russian Science Foundation (Grant No. 24-12-00167), and partially by the project FFUU-2024-0003 of the Institute of Spectroscopy of Russian Academy of Sciences.
\end{acknowledgments}

%

\end{document}